\documentclass[reprint,amsmath,amssymb,aps,pre,showpacs]{revtex4-1}

\usepackage{graphicx}
\usepackage{dcolumn}
\usepackage{bm}

 \usepackage{color}

\usepackage{amssymb}
\usepackage{graphicx}
\usepackage{amsbsy}

\newcommand\beq{\begin{equation}}
\newcommand\eeq{\end{equation}}
\newcommand\beqa{\begin{eqnarray}}
\newcommand\eeqa{\end{eqnarray}}
\newcommand{\nn}{\nonumber\\}

\newcommand{\HH}{H}
\newcommand{\ex}{\text{ex}}
\newcommand{\rr}{\mathbf{r}}
\newcommand{\nb}{\mathbf{n}}
\newcommand{\ww}{\bm{\omega}}
\newcommand{\wm}{\overline{\overline{\bm{\omega}}}_{2,i}}
\newcommand{\xxi}{\xi}

\newcommand{\ocite}{\cite}

\begin{document}



\title{Class of consistent fundamental-measure free energies for hard-sphere mixtures}


\author{Andr\'es Santos}
\email{andres@unex.es}
\homepage{http://www.unex.es/eweb/fisteor/andres/}
\affiliation{Departamento de F\'{\i}sica, Universidad de
Extremadura, Badajoz, E-06071, Spain}

\date{\today}

\begin{abstract}
In  fundamental-measure theories the bulk excess free-energy density of a hard-sphere fluid mixture is assumed to depend on the partial number densities $\{\rho_i\}$ only through the four scaled-particle-theory variables $\{\xxi_\alpha\}$, i.e., $\Phi(\{\rho_i\})\to\Phi(\{\xxi_\alpha\})$. By imposing consistency conditions, it is proven here that such a dependence must necessarily have the form $\Phi(\{\xxi_\alpha\})=-\xxi_0\ln(1-\xxi_3)+\Psi(y)\xxi_1\xxi_2/(1-\xxi_3)$, where $y\equiv {\xxi_2^2}/{12\pi \xxi_1 (1-\xxi_3)}$ is a scaled variable and  $\Psi(y)$  is an arbitrary dimensionless scaling function which can be determined from the free-energy density of the one-component system. Extension to the inhomogeneous case is achieved by standard replacements of the variables $\{\xxi_\alpha\}$ by the fundamental-measure (scalar, vector, and tensor) weighted densities $\{n_\alpha(\rr)\}$. Comparison with computer simulations shows the superiority of this bulk free energy over the White Bear one.
\end{abstract}

\date{\today}


\maketitle

\maketitle

\paragraph*{Introduction.}
Rosenfeld's fundamental-measure theory (FMT) \cite{R89,KR90}, together with its variants \cite{RSLT97,T00,RELK02,YW02,CMT02,YWXG04,M06,HR06},
is perhaps the most successful density-functional approach for the description of inhomogeneous hard-sphere (HS) systems. Recent reviews can be found in Refs.\ \ocite{TCM08,L10,R10}. FMT is based on the ansatz that the excess free-energy density $\Phi(\rr)$, which in general is a \emph{functional} of the partial number densities $\rho_i(\rr')$, can be approximated by a \emph{function} of a reduced number of \emph{weighted} densities \cite{P88} $\{n_\alpha(\rr)\}$, i.e.,
$\Phi[\rr,\{\rho_i(\rr')\}]\to\Phi(\{n_\alpha(\rr)\})$.  In order to choose a specific form for $\Phi(\{n_\alpha(\rr)\})$,   the \emph{bulk} free-energy density $\Phi(\{\rho_i\})\to\Phi(\{\xxi_\alpha\})$, where $\{\xxi_\alpha\}$ are  the so-called scaled-particle theory (SPT) variables, plays a crucial role. In Rosenfeld's original formulation \cite{R89,TCM08}, application of two thermodynamic consistency conditions [cf.\ Eqs.\ \eqref{thermo1} and  \eqref{thermo2} below] yields the SPT free energy. Later on, Roth \emph{et al.}\ \cite{RELK02} sacrificed consistency in favor of accuracy and proposed the so-called White Bear (WB) version, which is consistent with the successful Boubl\'ik-Mansoori-Carnahan-Starling-Leland (BMCSL) equation of state (EOS) \cite{B70,MCSL71} via Eq.\ \eqref{thermo1} but not via Eq.\ \eqref{thermo2}.
More recently, Hansen-Goos and Roth \cite{HR06,R10,HR06a} have proposed a modification of the WB functional (the so-called WB mark II) such that the inconsistency with Eq.\ \eqref{thermo2} disappears in the one-component limit.

The main aim of this Rapid Communication is to show that,  if the consistency conditions \eqref{thermo1} and  \eqref{thermo2} are complemented with a recently derived scaling relation \cite{S12}, the bulk free-energy density of the mixture is unambiguously expressed in terms of the bulk free energy of the \emph{one-component} system. If, in particular, the Carnahan-Starling (CS) expression \cite{CS69} is chosen as input, the output results for the bulk mixture are  more accurate than those resulting from the WB free energy.

\paragraph*{Bulk free-energy density.}
Let us consider a three-dimensional (additive) HS  fluid mixture  with (bulk) partial number densities $\rho_i$ and (bulk) total number density $\rho=\sum_i\rho_i$. The  moments of the size distribution are
$
M_\alpha\equiv \sum_i x_i\sigma_i^\alpha
$,
where $x_i\equiv\rho_i/\rho$ and $\sigma_i$ are the mole fraction and diameter, respectively, of spheres of component $i$.

Let  $\Phi(\{\rho_i\})$ be the bulk excess free-energy density, in units of $k_BT$, where $k_B$ is the Boltzmann constant and $T$ is the absolute temperature. The free energy is said to be ``truncatable'' \cite{GKM82,SWC01,S02} if $\Phi(\{\rho_i\})$ depends on $\{\rho_i\}$ only through  the total packing fraction $\eta\equiv(\pi/6)\rho M_3$ and  a \emph{finite} number $K$ of moments $\{M_1, M_2,\ldots, M_K\}$.  According to FMT arguments, the relevant number of moments for three-dimensional systems is $K=3$.
In that case, instead of the set of variables $\{\eta;M_1,M_2,M_3\}$, one can alternatively take the four SPT variables
$\xxi_0=\rho$,
$\xxi_1=\frac{1}{2}\rho M_1$,
$\xxi_2=\pi \rho M_2$, and
$\xxi_3=\frac{\pi}{6}\rho M_3$.

By requiring consistency with the limit where one of the components is made of point particles, it has been proven in Ref.\ \ocite{S12} that a truncatable free energy (with $K=3$)  must necessarily have the scaling property
\beq
\Phi(\{\xxi_\alpha\})=-\xxi_0\ln(1-\xxi_3)+{\xxi_1\xxi_2}\HH(\xxi_3,z),
\label{1}
\eeq
where we have called
$
z\equiv {\xxi_2^2}/{12\pi \xxi_1 }
$
and the dimensionless scaling function $H(\xxi_3,z)$ remains undetermined.
Equation \eqref{1} imposes an important constraint on the functional dependence of $\Phi$ on the four SPT variables $\{\xxi_\alpha\}$ since the unknown function $\HH$ depends on two variables only \cite{S12}. The main goal of this Rapid Communication is to prove that an extra consistency condition further constrains $\Phi$ to an unknown function of one variable only.

First, we consider the standard thermodynamic relation
\beq
\beta p=\rho+\rho^2\frac{\partial}{\partial\rho}\frac{\Phi}{\rho}=\xxi_0-\Phi+\sum_{\alpha=0}^3\xxi_\alpha\frac{\partial\Phi}{\partial \xxi_\alpha},
\label{thermo1}
\eeq
where $p$ is the pressure and $\beta\equiv 1/k_BT$. An independent SPT condition  is \cite{RELK02,RFHL60}
\beq
\beta p=\lim_{\sigma_i\to\infty}\frac{\beta\mu_i^\ex}{{\pi}\sigma_i^3/{6}}=\frac{\partial\Phi}{\partial \xxi_3},
\label{thermo2}
\eeq
where $\mu_i^\ex$ is the excess chemical potential of particles of diameter $\sigma_i$.
The first equality of Eq.\ \eqref{thermo2} is related to the reversible work needed to create a cavity large enough to accommodate a particle of infinite diameter \cite{RELK02}. Note that the second equalities in Eqs.\ \eqref{thermo1} and \eqref{thermo2} are valid \emph{only} if $\Phi(\{\rho_i\}) \to \Phi(\{\xxi_\alpha\})$.
Internal consistency between Eqs.\ \eqref{thermo1} and \ref{thermo2} implies
\beq
(1-\xxi_3)\frac{\partial\Phi}{\partial \xxi_3}=\xxi_0-\Phi+\sum_{\alpha=0}^2\xxi_\alpha\frac{\partial\Phi}{\partial \xxi_\alpha}.
\label{consist}
\eeq
Equation \eqref{consist} is independent of the scaling relation \eqref{1}. Inserting the latter into the former one obtains the linear partial differential equation
$\HH(\xxi_3,z)=(1-\xxi_3){\partial_{\xxi_3} \HH(\xxi_3,z)}-z{\partial_z \HH(\xxi_3,z)}$,
whose solution is
$
\HH(\xxi_3,z)=\frac{1}{1-\xxi_3}\Psi\left(\frac{z}{1-\xxi_3}\right)
$,
where the function $\Psi(y)$ remains undetermined.

Therefore, if the (bulk) excess free-energy density of the HS mixture depends on the partial densities only through the four SPT variables, then it must necessarily have the simple scaling form
\beq
\Phi(\{\xxi_\alpha\})=-\xxi_0\ln(1-\xxi_3)+\frac{\xxi_1\xxi_2}{1-\xxi_3}\Psi(y),
\label{3}
\eeq
where we have called
\beq
y\equiv \frac{\xxi_2^2}{12\pi \xxi_1 (1-\xxi_3)}.
\label{y}
\eeq
The pressure follows from either Eq.\ \eqref{thermo1} or \eqref{thermo2} as
\beq
\beta p=\frac{\xxi_0}{1-\xxi_3}+\frac{\xxi_1\xxi_2}{(1-\xxi_3)^2}\left[\Psi(y)+y\Psi'(y)\right],
\label{press}
\eeq
where $\Psi'(y)=d \Psi(y)/dy$.
Agreement with the exact second and third virial coefficients simply requires $\Psi(y)=1+\frac{1}{2}y+\mathcal{O}(y^2)$.
The excess chemical potential of component $i$ is obtained from Eq.\ \eqref{3} as
\beqa
\beta\mu_i^\ex&=&\frac{\partial \Phi}{\partial\rho_i}=-\ln(1-\xxi_3)+\frac{\xxi_2}{1-\xxi_3}\left[\Psi(y)-y\Psi'(y)\right]\frac{\sigma_i}{2}\nn
&&+\frac{\xxi_1}{1-\xxi_3}\left[\Psi(y)+2y\Psi'(y)\right]\pi\sigma_i^2+\beta p\frac{\pi\sigma_i^3}{6}.
\label{mu}
\eeqa

The unknown function $\Psi(y)$ can be univocally derived from the one-component free-energy density. Setting $\sigma_i=\sigma$, Eq.\ \eqref{3} reduces to
\beq
\phi_s(\eta)=-\ln(1-\eta)+3\frac{\eta}{1-\eta}\Psi\left(\frac{\eta}{1-\eta}\right),
\label{4}
\eeq
where $\phi_s(\eta)=\Phi(\eta)/\rho$ is the (bulk) excess free energy \emph{per particle} of the one-component system. {}From Eq.\ \eqref{4} one gets
\beq
\Psi(y)=\frac{1}{3y}\left[\phi_s\left(\frac{y}{1+y}\right)-\ln(1+y)\right].
\label{5}
\eeq
The combination of Eqs.\ \eqref{3}, \eqref{y}, and \eqref{5} constitutes the main result of this Rapid Communication. It shows that, given the free energy of the pure HS system, the free energy of the HS mixture (if assumed to be truncatable with $K=3$) is unambiguously known. In a more explicit form,
\beqa
\Phi(\{\xxi_\alpha\})&=&-\xxi_0\ln(1-\xxi_3)+4\pi\frac{\xxi_1^2}{\xxi_2}\nn
&&\times\left[\phi_s\left(\frac{y}{1+y}\right)-\ln(1+y)\right].
\label{6}
\eeqa
Analogously, Eq.\ \eqref{press} yields
\beq
\beta p=\frac{\xxi_0}{1-\xxi_3}+4\pi\frac{\xxi_1^2}{\xxi_2(1-\xxi_3)}\left[\frac{Z_s\left(\frac{y}{1+y}\right)}{1+y}-1\right],
\label{press2}
\eeq
where $Z_s(\eta)$ is the compressibility factor $\beta p/\rho$ of the one-component system.
According to Eqs.\ \eqref{6} and \eqref{press2}, the thermodynamic properties of a HS mixture of total packing fraction $\eta$ can be expressed in terms of those of a pure HS fluid with an \emph{effective} packing fraction $\eta_{\text{eff}}=y/(1+y)=[1+12\pi(1-\xxi_3)\xxi_1/\xxi_2^2]^{-1}=[1+(1-\eta)M_1M_3/\eta M_2^2]^{-1}\leq \eta$.
This contrasts with other approaches \cite{HYS08} in which the reference one-component fluid has the same packing fraction ($\eta_{\text{eff}}=\eta$) as the mixture.

It is worth noting that, while Eq.\ \eqref{press2} reproduces the exact second and third virial coefficients, it predicts a fourth virial coefficient given by $B_4=\left({\pi}/{6}\right)^3\left[M_3^3+9M_3 M_2^3+9M_1M_2M_3^2+(b_4-19){M_2^5}/{M_1}\right]$, where $b_4$ is the reduced fourth virial coefficient of the one-component fluid. This expression is not consistent with a polynomial dependence on the mole fractions, $B_4=\sum_{i,j,k,\ell}x_i x_j x_k x_\ell B_{ijk\ell}$, except in the SPT case ($b_4=19$). This is a consequence of the approximate character of the FMT (or truncatability) ansatz  at the level of the fourth virial coefficient, as already pointed out by Blaak \cite{B98}.

A milder proposal consists of making the formal change $\Psi(y)\to A_0(\xxi_3)+A_1(\xxi_3)y$ in Eq.\ \eqref{3}, where the functions $A_0(\xxi_3)$ and $A_1(\xxi_3)$ satisfy the differential equation $(1-\xxi_3)A_0'(\xxi_3)+\xxi_3A_1'(\xxi_3)=0$ with the initial conditions $A_0(0)=1$, $A_1(0)=\frac{1}{2}$. Although, in the general case, this breaks the consistency condition \eqref{consist}, the latter is verified in the one-component limit $\xxi_2^2/12\pi\xxi_1\xxi_3\to 1$. Given a desired compressibility factor $Z_s(\eta)$, the  differential equation is closed with the algebraic relation $(1-\xxi_3)A_0(\xxi_3)+2\xxi_3A_1(\xxi_3)=(1-\xxi_3)^2[(1-\xxi_3)Z_s(\xxi_3)-1]/3\xxi_3$.
Choosing for $Z_s(\eta)$ the CS EOS yields the WB mark II free energy, which was proposed in Refs.\ \cite{HR06,HR06a} by a different method.

\paragraph*{The inhomogeneous case.}
Let us start by writing $\Psi(y)=1+\frac{1}{2}y+\Lambda(y)$, where $\Lambda(y)=\mathcal{O}(y^2)$, so that Eq.\ \eqref{3} becomes
\beq
\Phi=\Phi_1+\Phi_2+\Phi_3+\Phi_2 \Lambda\left(\frac{2\Phi_3}{\Phi_2}\right),
\label{3alt}
\eeq
where $\Phi_1\equiv -\xxi_0\ln(1-\xxi_3)$, $\Phi_2\equiv {\xxi_1\xxi_2}/(1-\xxi_3)$, and $\Phi_3\equiv \xxi_2^3/24\pi (1-\xxi_3)^2$. The SPT free energy corresponds to the \emph{linear} approximation $\Lambda(y)=0$.

In Rosenfeld's original FMT \cite{R89}, the excess free-energy \emph{functional} $\Phi[\rr,\{\rho_i(\rr')\}]$ in inhomogeneous situations can be constructed from the SPT bulk quantity [Eq.\ \eqref{3alt} with $\Lambda(y)=0$] by following two basic steps. First, the four bulk SPT variables $\{\xxi_0,\xxi_1,\xxi_2,\xxi_3\}$ are replaced by the weighted densities
\beq
n_\alpha(\rr)=\sum_i\int d\rr'\, \rho_i(\rr-\rr')\omega_{\alpha,i}(\rr'),
\label{I1}
\eeq
where the (scalar) weight functions are
$
\omega_{0,i}(\rr)={\omega_{2,i}(\rr)}/{\pi\sigma_i^2}$, $\omega_{1,i}(\rr)={\omega_{2,i}(\rr)}/{2\pi\sigma_i}$,
$\omega_{2,i}(\rr)=\delta(r-\sigma_i/2)$, and $\omega_{3,i}(\rr)=\Theta(\sigma_i/2-r)$.
Second, to those four scalar weighted densities  two vector weighted densities $\nb_1(\rr)$ and $\nb_2(\rr)$ are added. They are defined similarly to Eq.\ \eqref{I1}, but with the vector weight functions
$\ww_{1,i}(\rr)=\omega_{1,i}(\rr){\rr}/{r}$ and $\ww_{2,i}(\rr)=\omega_{2,i}(\rr){\rr}/{r}$.
The two vector densities $\nb_1$ and $\nb_2$ vanish in the bulk and so they are absent in the bulk free energy. A simple way of including $\nb_1$ and $\nb_2$ in $\Phi(\rr)$ and preserving the exact low-density behavior consists of making the formal changes $\xxi_1\xxi_2\to n_1 n_2-\nb_1\cdot\nb_2$ and $\xxi_2^3\to n_2(n_2^2-3\nb_2\cdot\nb_2)$ in $\Phi_2$ and $\Phi_3$, respectively. Therefore, Rosenfeld's functional is given by the right-hand side of Eq.\ \eqref{3alt} with $\Lambda(y)=0$ and
\beq
\Phi_1(\{n_\alpha(\rr)\})=-n_0(\rr)\ln[1-n_3(\rr)],
\label{Phi1}
\eeq
\beq
\Phi_2(\{n_\alpha(\rr)\})=\frac{n_1(\rr)n_2(\rr)-\nb_1(\rr)\cdot\nb_2(\rr)}{1-n_3(\rr)},
\label{Phi2}
\eeq
\beq
\Phi_3(\{n_\alpha(\rr)\})=\frac{n_2(\rr)[n_2^2(\rr)-3\nb_2(\rr)\cdot \nb_2(\rr)]}{24\pi[1-n_3(\rr)]^2}.
\label{Phi3}
\eeq

In order to recover the exact one-dimensional (1D) functional for one-component systems (``dimensional crossover''), Tarazona \cite{T00} proposed to modify  Eq.\ \eqref{Phi3} as
\beq
\Phi_3(\{n_\alpha\})=\frac{n_2(n_2^2-3\nb_2\cdot \nb_2)+\frac{9}{2}[\nb_2\cdot\mathsf{n}_2\cdot\nb_2-\text{Tr}(\mathsf{n}_2^3)]}{24\pi (1-n_3)^2},
\label{I8}
\eeq
where the additional tensor weighted density $\mathsf{n}_2(\rr)$ is defined by Eq.\ \eqref{I1} with the tensor weight function \cite{RELK02,T00}
$\wm(\rr)=\omega_{2,i}(\rr)\left({\rr\rr}/{r^2}-{\mathsf{I}}/{3}\right)$,
$\mathsf{I}$ being the unit tensor.
In the special case of a one-component system ($\sigma_i=\sigma$) confined to a 1D geometry, i.e., $\rho_i(\rr)=x_i\rho^{(\text{1D})}\delta(x)\delta(y)$, one has $n_2=\rho^{(\text{1D})}\sigma\Theta(\sigma/2-t)/\sqrt{(\sigma/2)^2-t^2}$, $\nb_2/n_2=(2/\sigma)\mathbf{t}$, and $\mathsf{n}_2/n_2=(4/\sigma^2)\left(\mathbf{t}\mathbf{t}-t^2\widehat{\mathbf{z}}\widehat{\mathbf{z}}\right)+
\widehat{\mathbf{z}}\widehat{\mathbf{z}}-\frac{1}{3}\mathsf{I}$, where $t\equiv\sqrt{x^2+y^2}$ and $\mathbf{t}\equiv x\widehat{\mathbf{x}}+y\widehat{\mathbf{y}}$. Insertion into Eq.\ \eqref{I8} yields $\Phi_3=0$ and spatial integration over the remaining two terms $\Phi_1+\Phi_2$ provides the exact 1D free energy \cite{RSLT97}.

In the case of the general class of consistent bulk free-energy densities \eqref{3alt}, it seems quite natural to construct the corresponding class of functionals by applying the replacements \eqref{Phi1} and \eqref{Phi2}, together with either Eq.\ \eqref{Phi3} (vector densities) or Eq.\ \eqref{I8} (tensor densities). In order to specify a particular functional one only needs to choose a thermodynamic description for the pure fluid as input and then obtain the function $\Psi(y)=1+\frac{1}{2}y+\Lambda(y)$ via Eq.\ \eqref{4} or, equivalently, Eq.\ \eqref{5}.
Since $\Lambda(y)=\mathcal{O}(y^2)$ and $\Phi_3$, as given by Eq.\ \eqref{I8}, vanishes in the one-component 1D limit, it is obvious that  the whole class of free-energy densities \eqref{3alt} becomes exact in that limit.
On the other hand, the presence of the nonlinear term $\Lambda(y)$ in Eq.\ \eqref{3alt} induces in the associated direct correlation function a spurious simple pole at $r=0$ \cite{L12}.

\paragraph*{Application to bulk properties.}
As an application of our scheme, let us assume that the bulk excess free energy per particle of the pure fluid has the form
\beqa
\phi_s(\eta)&=&-\ln (1-\eta)+3\eta\frac{1-\eta/2}{(1-\eta)^2}\nn
&&+\lambda\left[\eta\frac{1-3\eta/2}{(1-\eta)^2}+\ln(1-\eta)\right].
\label{lambda}
\eeqa
The associated EOS of the pure HS fluid is
\beq
Z_s(\eta)=\frac{1+\eta+\eta^2-\lambda\eta^3}{(1-\eta)^3}.
\label{8}
\eeq
The forms \eqref{lambda}  and \eqref{8} encompass the virial-route Percus-Yevick (PY)  ($\lambda=3$), the compressibility-route PY (or SPT) ($\lambda=0$), and the CS  \cite{CS69} ($\lambda=1$) EOSs. In general, the coefficient $\lambda$ is related to the reduced fourth virial coefficient  by $b_4=19-\lambda$. Insertion of Eq.\ \eqref{lambda} into Eq.\ \eqref{5} yields
\beq
\Psi(y)=1+\frac{1}{2}y+\frac{\lambda}{3}\left[1-\frac{1}{2}y-\frac{\ln(1+y)}{y}\right].
\label{7}
\eeq
In the SPT case ($\lambda=0$), $\Psi(y)$ is simply approximated by the exact expansion in powers of $y$ truncated after the linear term. In general,  Eq.\ \eqref{7} gives
$\Psi(y)=1+\frac{1}{2}y-(\lambda/9) y^2+\mathcal{O}(y^3)$.
According to Eq.\ \eqref{3}, the \emph{unique} free-energy density of the mixture consistent with the one-component expression \eqref{lambda} is
\beqa
\Phi(\{\xxi_\alpha\})&=&-\xxi_0\ln(1-\xxi_3)+\left(1+\frac{\lambda}{3}\right)\frac{\xxi_1\xxi_2}{1-\xxi_3}
\nn
&&+\left(1-\frac{\lambda}{3}\right)\frac{\xxi_2^3}{24\pi (1-\xxi_3)^2}\nn
&&-4\pi {\lambda}\frac{\xxi_1^2}{\xxi_2}\ln\left[1+\frac{\xxi_2^2}{12\pi \xxi_1 (1-\xxi_3)}\right].
\label{9}
\eeqa
The associated EOS can be obtained from either Eq.\ \eqref{thermo1} or Eq.\ \eqref{thermo2}. The result is
\beqa
\beta p&=&\frac{\xxi_0}{1-\xxi_3}+\left(1+\frac{\lambda}{3}\right)\frac{\xxi_1\xxi_2}{(1-\xxi_3)^2}
\nn&&
+\left(1-\frac{\lambda}{3}\right)\frac{\xxi_2^3}{12\pi (1-\xxi_3)^3}\nn &&
-\frac{4\pi\lambda}{1-\xxi_3}\frac{\xxi_1^2 \xxi_2}{\xxi_2^2+12\pi \xxi_1(1-\xxi_3)}.
\label{10}
\eeqa
It is easy to check that Eqs.\ \eqref{9} and \eqref{10} reduce to Eqs.\ \eqref{lambda} and \eqref{8}, respectively, in the one-component limit $\sigma_i\to\sigma$.

If we take as input the SPT thermodynamic description for the one-component fluid, i.e., Eq.\ \eqref{lambda} with  $\lambda=0$, then the output free energy for the mixture given by Eq.\ \eqref{9}  coincides with the true SPT result for mixtures. This is an expected result since, as is well known, the SPT is thermodynamically consistent with Eqs.\ \eqref{thermo1} and \eqref{thermo2}. On the other hand, the SPT EOS is known to overestimate the pressure of the HS fluid. A much better description is provided by the CS description, i.e., by Eqs.\ \eqref{lambda} and \eqref{8} with $\lambda=1$. Taking this as input, the consistent extension to mixtures is provided by Eqs.\ \eqref{9} and \eqref{10} with $\lambda=1$. The result differs from the most popular extension of the CS equation to mixtures, namely the BMCSL theory \cite{B70,MCSL71}. The BMCSL EOS is
\beq
\beta p=\frac{\xxi_0}{1-\xxi_3}+\frac{\xxi_1\xxi_2}{(1-\xxi_3)^2}
+\frac{\xxi_2^3(3-\xxi_3)}{36\pi (1-\xxi_3)^3}.
\label{BMCSL1}
\eeq
{}According to the thermodynamic relation \eqref{thermo1}, the free-energy density corresponding to Eq.\ \eqref{BMCSL1} is
\beqa
\Phi(\{\xxi_\alpha\})&=&-\xxi_0\ln(1-\xxi_3)+\frac{\xxi_1\xxi_2}{1-\xxi_3}
\nn
&&+\frac{\xxi_2^3}{36\pi \xxi_3^2}\left[\frac{\xxi_3}{(1-\xxi_3)^2}+\ln(1-\xxi_3)\right].
\label{BMCSL2}
\eeqa
On the other hand, if Eq.\ \eqref{thermo2} is used instead, the result is
\beq
\Phi(\{\xxi_\alpha\})=-\xxi_0\ln(1-\xxi_3)+\frac{\xxi_1\xxi_2\xxi_3}{1-\xxi_3}
+\frac{\xxi_2^3\xxi_3(3-2\xxi_3)}{36\pi (1-\xxi_3)^2}.
\label{BMCSL3}
\eeq
The difference between Eqs.\ \eqref{BMCSL2} and \eqref{BMCSL3} reflects that, in contrast to Eq.\ \eqref{10}, the BMCSL EOS \eqref{BMCSL1} is inconsistent with respect to the simultaneous verification of the thermodynamic relations \eqref{thermo1} and \eqref{thermo2}. The free-energy density \eqref{BMCSL2} was the one used in Ref.\ \ocite{RELK02} to construct the WB version of the FMT.

The alternate EOS \eqref{10}, as well as the BMCSL,  PY, and SPT EOSs, share the property that, at a given total packing fraction, the compressibility factor $\beta p/\rho$ depends on the size distribution only through the dimensionless combinations of moments $\gamma_1\equiv M_1 M_2/M_3$ and $\gamma_2\equiv M_2^3/M_3^2$. On the other hand, while the dependence of the PY, SPT, and BMCSL EOSs on both $\gamma_1$ and $\gamma_2$ is \emph{linear}, the last term on the right-hand side of Eq.\ \eqref{10} introduces a nonlinear dependence.

\begin{table}
\caption{Difference between theoretical and simulation values \cite{BMLS96} for the compressibility factor $\beta p/\rho$ and the (reduced) chemical potentials $\beta\mu_1$ and $\beta\mu_2$ in a binary mixture with a total packing fraction $\eta=0.49$, a size ratio $\sigma_2/\sigma_1=0.3$, and several mole fractions $x_1$.
\label{tab}}
\begin{ruledtabular}
\begin{tabular} {ccccccccc}
 &\multicolumn{2}{c}{$\Delta (\beta p/\rho)$}&&\multicolumn{2}{c}{$\Delta(\beta\mu_1)$}&&\multicolumn{2}{c}{$\Delta(\beta\mu_2)$}\\
 \cline{2-3} \cline{5-6}\cline{8-9}
$x_1$&Eq.\ \protect\eqref{9}&Eq.\ \protect\eqref{BMCSL2}&&Eq.\ \protect\eqref{9}&Eq.\ \protect\eqref{BMCSL2}&&Eq.\ \protect\eqref{9}&Eq.\ \protect\eqref{BMCSL2}\\
\hline
$\frac{1}{16}$ &$-0.02$&$-0.08$&&$-0.1$&$-0.8$&&$-0.01$&$-0.04$\\
$\frac{1}{8}$ &$-0.01$& $-0.07$&&$-0.05$&$-0.4$&&$-0.002$&$-0.03$\\
$\frac{1}{4}$ &$-0.007$& $-0.05$&&$-0.0002$&$-0.2$&&$0.001$&$-0.02$\\
$\frac{1}{2}$ &$-0.01$& $-0.04$&&$-0.006$&$-0.07$&&$0.004$&$-0.01$\\
$\frac{3}{4}$ &$-0.02$& $-0.04$&&$-0.03$&$-0.05$&&$0.005$&$-0.01$\\
\end{tabular}
\end{ruledtabular}
\end{table}

As a test of the superiority of this free-energy density, Eq.\ \eqref{9} with $\lambda=1$, over the BMCSL one, Eq.\ \eqref{BMCSL2}, Table \ref{tab} compares the respective deviations of the theoretical values with respect to the simulation ones \cite{BMLS96} for the compressibility factor $\beta p/\rho$ and the (reduced) chemical potentials $\beta\mu_1$ and $\beta\mu_2$. The system chosen is the one with the highest packing fraction ($\eta=0.49$) and the largest size disparity ($\sigma_2/\sigma_1=0.3$) considered in Ref.\ \ocite{BMLS96}. We observe that the deviations are strongly reduced by the use of the consistent free energy \eqref{9}.
Comparison with Table I of Ref.\ \cite{HR06a} shows that the WB mark II free energy predicts even better values for the pressure. However, Eq.\ \eqref{9} predicts more accurate values of $\mu_1$ for $x_1=\frac{1}{4}$ and of $\mu_2$ for $x_1=\frac{1}{8},\frac{1}{4},\frac{1}{2}$. For the largest mole fraction ($x_1=\frac{3}{4}$), Eq.\ \eqref{9} and the WB mark II free energy give practically the same results for $p$, $\mu_1$, and $\mu_2$.

It is interesting to remark that Eq.\ \eqref{9} with $\lambda=1$ does predict a (metastable) demixing transition, in contrast to the SPT [Eq.\ \eqref{9} with $\lambda=0$] and the BMCSL [Eq.\ \eqref{BMCSL2}] descriptions. In particular, the predicted coordinates of the critical point are $(x_{1c},\eta_c,\beta p_c\sigma_1^3)=(0.0022,0.686,7305)$ and $(0.000\,32,0.594,19\,404)$  for the size ratios $\sigma_2/\sigma_1=\frac{1}{10}$ and $\frac{1}{20}$, respectively.

\paragraph*{Conclusions.}
To sum up, once the FMT (or truncatability) ansatz for the bulk free-energy density  is assumed, basic consistency conditions impose that it must necessarily have the functional form expressed by Eq.\ \eqref{3}, where the scaled variable $y$ is given by Eq.\ \eqref{y} and the dimensionless scaling function $\Psi(y)$ remains arbitrary, except for the low-density requirement $\Psi(y)=1+\frac{1}{2}y+\mathcal{O}(y^2)$. This arbitrariness, however, can be exploited to seek consistency with any desired EOS in the one-component case via Eqs.\ \eqref{4} or \eqref{5}. Thus, Eq.\ \eqref{3} represents a \emph{class} of consistent FMT free energies, parametrized by the one-component function. This class includes the SPT free energy [Eq.\ \eqref{9} with $\lambda=0$] as a particular case. However, the WB free energy \eqref{BMCSL2} does not belong to the class. In fact, the unique member of the class consistent with the CS EOS [Eq.\ \eqref{9} with $\lambda=1$] turns out to be much more accurate than the WB free energy, as shown by Table \ref{tab}. Of course, other choices are possible and it can be reasonably expected that the more accurate the  one-component input, the better the  multicomponent output.

While most of the Rapid Communication has dealt with the bulk free energy, the extension of the results to confined geometries and/or to external potentials is straightforwardly given by Eq.\ \eqref{3alt} complemented by Eqs.\ \eqref{Phi1} and \eqref{Phi2} plus Eq.\ \eqref{Phi3} (including vector weighted densities) or Eq.\ \eqref{I8} (including tensor weighted densities).

As any other approximation, the proposal presented here has strengths and weaknesses. Weak points are that, except for the genuine SPT free energy, the virial coefficients beyond the third one do not depend polynomially on the mole fractions and the direct correlation function stemming from the inhomogeneous functional possesses a pole at $r=0$. These shortcomings are reflections of the approximate character of the FMT ansatz $\Phi(\{\rho_i\})\to\Phi(\{\xxi_\alpha\})$. On the other hand, strong features are the internal consistency with Eqs.\ \eqref{1} and \eqref{consist}, the flexibility to accommodate any desired one-component thermodynamic description, and the accuracy of the bulk free energy when an accurate one-component theory is chosen as input. Moreover, preliminary calculations for confined fluids provide encouraging results \cite{GWS12}.

\paragraph*{Acknowledgments.} I am grateful to J.\ Kolafa, M.\ L\'opez de Haro, J.\ F.\ Lutsko, A.\ Malijevsk\'y, Al.\ Malijevsk\'y, R.\ Roth, M.\ Schmidt, P.\ Tarazona, and J.\ A.\ White for insightful and stimulating comments. Financial support from the Spanish Government through Grant No. FIS2010-16587 and from the Junta de Extremadura (Spain) through Grant No.\ GR10158 (partially financed by FEDER funds) is  acknowledged.

\bibliographystyle{apsrev}
\bibliography{D:/Dropbox/Public/bib_files/liquid}
\end{document}